# Scattering, weak localization and Shubnikov-de Haas oscillation in high carrier density AlInN/GaN heterostructures

Leizhi Wang,[1] Ming Yin,[2] Asif Khan,[3] Sakib Muhtadi,[3] Fatima Asif,[3] Eun Sang Choi,[4] and Timir Datta[1*]

[1]*Department of Physics and Astronomy, University of South Carolina, Columbia, South Carolina 29208, USA*

[2]*Department of Physics and Engineering, Benedict College, Columbia, South Carolina, 29204, USA*

[3]*Department of Electrical Engineering, University of South Carolina, Columbia, South Carolina 29208, USA*

[4]*National High Magnetic Field Laboratory, Tallahassee, Florida, 32310, USA*

We provide the first observation of weak localization in high carrier density ($\sim 2 \times 10^{13} cm^{-2}$) two-dimensional electron gas in AlInN/GaN heterostructures; at low temperatures and low fields the conductivity increases with increasing magnetic field. Weak localization is further confirmed by the *lnT* dependence of the zero-field conductivity and angle dependence of magnetoresistance. The inelastic scattering rate is linearly proportional to temperature, $\tau_i^{-1} \propto T$, demonstrating that electron-electron scattering is the principal phase breaking mechanism. Shubnikov-de Haas (SdH) oscillations at high magnetic fields are also observed. From the temperature dependent amplitude of SdH oscillation and Dingle plot, the effective mass of electron is extracted to be $0.2327 m_e$; in addition the quantum lifetime is smaller than transport time from Hall measurement, indicating small angle scattering such as from remote ionized impurities is dominant. Above 20 K, the scattering changes from acoustic phonon to optical phonon scattering, resulting in a rapid decrease in carrier mobility with increasing temperature.

## I. INTRODUCTION

GaN based III-V semiconductors, in particular AlInN/GaN heterostructures, have attracted great interest due to potential applications in high power and high frequency electronics as well as high electron mobility transistors (HEMT). Two-dimensional electron gas (2DEG) in GaN heterostructure arises from the polarization induced surface charge instead of doping, which can reduce the impurity scatterings. In contrast to AlGaN, there are several advantages with AlInN as the barrier layer. First, one can achieve lattice matching to GaN by tuning the composition between AlN and InN, which may greatly improve the crystal quality and carrier mobility. Furthermore, AlInN barrier can provide higher carrier densities due to the larger spontaneous polarization [1,2]. Additionally, high thermal and chemical stabilities of AlInN allow device operation above 1000 °C [3]. Hence AlInN/GaN systems are particularly suitable for high power and high frequency electronics.

In order to improve the quality of heterostructures, it is necessary to understand the scattering mechanisms. Transport measurements are effective tools for elucidating scattering process and can provide information about Fermi surface, effective mass, quantum dephasing and relaxation times [4-8]. However, scattering mechanisms and quantum transports in this new AlInN/GaN system have not yet been widely investigated.

---

* Corresponding author: datta@sc.edu

At low temperatures phonon scattering is suppressed, which induces long mean free path and coherence length. In this regime, due to the constructive quantum interference, the carrier has an enhanced probability to be scattered back to the origin along the closed loop in opposite directions. This coherent back scattering leads to weak localization [5,9-11] and an increase in resistance compared with the classical Drude value. The application of a magnetic field breaks the time reversal symmetry and destroys the interference. Hence localization probability is decreased which produces a counterintuitive increase in the magnetoconductivity with increasing magnetic field. The elastic scattering time and the inelastic phase breaking time can be readily obtained from the magnetoconductivity.

At high magnetic fields, the band structure becomes quantized into Landau levels and carriers in the interior region execute cyclotron motion. But at the boundaries, orbital motion is disrupted, the carriers get scattered forward along the edge, leading to a super flow of current and increase in conductance. With increasing magnetic field, the resistance oscillates periodically as a function of $1/B$, known as Shubnikov-de Haas (SdH) oscillation [6,8,12,13], which is useful to analyze the effective mass and quantum scattering mechanism.

In this article, we investigate the scattering mechanisms by electrical and magneto transport measurements in AlInN/GaN heterostructures. We report the first observation of weak localization effect in this new system. Our analysis of data below 20 K demonstrates that the inelastic scattering rate is linearly proportional to temperature, $\tau_i^{-1} \propto T$, indicating electron-electron scattering to be the dominant phase breaking mechanism. Weak localization is further confirmed by the $lnT$ dependence of the zero-field conductivity. We also report on Shubnikov-de Haas oscillations. From the temperature dependent amplitude of SdH and Dingle plot, we extract the effective mass of electron and quantum lifetime. Above 20 K, both weak localization and SdH oscillations are suppressed. In this high temperature regime, the scattering changes from acoustic phonon to optical phonon scattering with the increasing temperature.

## II. EXPERIMENT

The $Al_{0.83}In_{0.17}N$/GaN epilayer structures were grown on sapphire substrates by standard metal-organic chemical vapor deposition (MOCVD) process [14]. A SEM image of the cross section view of such a heterostructure is provided in Fig. 1(a). It clearly shows a ~200 nm AlN buffer layer followed by ~2.2 μm undoped GaN as channel layer, ~1 nm AlN spacer and ~7 nm AlInN barrier layer with In composition of 17%. The Hall bar mesas were defined by photolithography and followed by an inductive coupled plasma etching using $Cl_2/BCl_3$, as shown in the schematic diagram of Fig. 1(b). Six Ohmic contacts were formed by evaporating Ti/Al/Ti/Au (40/120/40/80 nm) metal stacks using e-beam deposition and then annealed at 850°C for 30 sec in $N_2$ ambient in order to diffuse Al to the 2DEG region. The transport measurements were conducted in an 18/20 Tesla Superconducting Magnet (SCM2) with $^3$He insert at the NHMFL. The input current with 1 μA at 17.97 Hz was applied by a lock-in amplifier (SR830 DSP).

## III. RESULTS AND DISCUSSION
### A. Temperature dependent electrical transport

The typical temperature dependence of sheet resistance, $R_\square$ is shown in Fig. 1(b). Generally the sheet resistance increases with increasing temperature above 20 K, consistent with metallic-like transport. The variation of the Hall carrier density as a function of temperature is shown in Fig. 1(c). Although the density increases with increasing temperature, the change is very small. This is because the band gap is too large: $E_g \sim$ 4 eV [15,16]. In the regime of interest (280 mK~280 K), $k_B T \ll E_g$; thus the carrier density is nearly temperature independent.

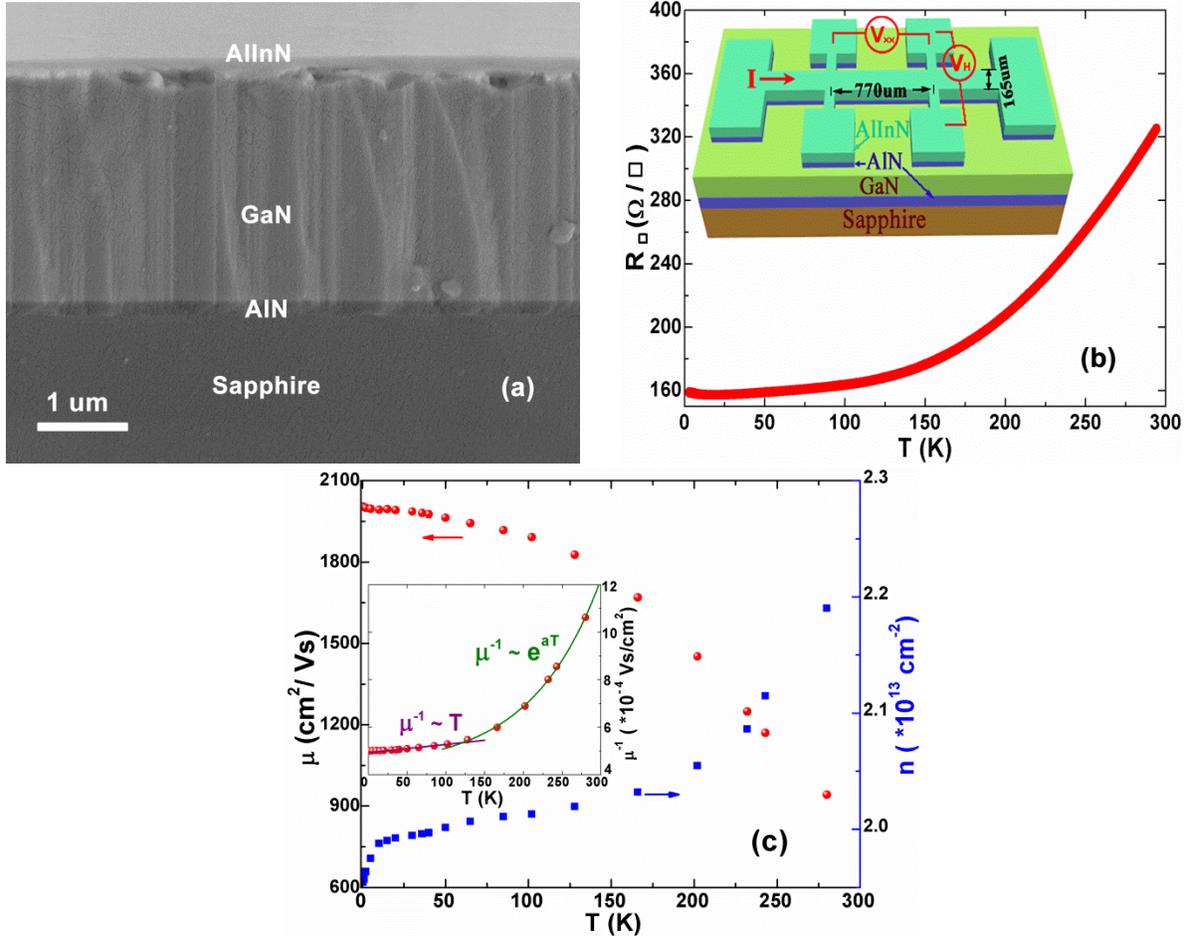

FIG. 1. (a) Cross-section SEM image of AlInN/GaN heterostructure. (b) Temperature dependent sheet resistance, inset is the schematic diagram of structure. (c) Carrier density and Hall mobility as a function of temperature.

In contrast, the Hall mobility decreases with increasing temperature, and the decreasing rate exhibits an interesting variation with temperature. Below 20 K, the mobility is weakly temperature dependent. It decreases slightly with increasing temperature, mirroring that of the sheet resistance. As will be discussed later, these behaviors may arise from impurity scattering or surface roughness as well as electron-electron scattering. Between 20 K and 100 K, the temperature dependence is more pronounced. As shown in the inset of Fig. 1(c), the inverse mobility is directly proportional to temperature, $\mu^{-1} \propto T$, indicating that acoustic phonon scattering is dominant [2,17,18]. Such temperature dependence has been widely observed in AlGaN/GaN heterostructures. At higher temperatures, above 100 K, the mobility decreases even faster and an exponential dependence $\mu^{-1} \propto e^{aT}$ describes the data very well. This is indicative of highly temperature dependent scattering of carriers. In the 2DEG literature, such exponentially temperature dependent mobility at high temperatures has been attributed to polar optical phonon scattering [2,19]. Consequently the mobility is reduced with increasing temperature throughout the temperature range we measured.

## B. Shubnikov-de Haas oscillation

The longitudinal resistance $R_{xx}$ as a function of applied magnetic field $B$ up to 18 T at different temperatures is shown in Fig. 2(a). As the magnetic field is increased, Shubnikov-de Haas (SdH) oscillations appear. The peaks of the oscillations are pronounced at low temperatures and damped with increasing temperature. This effect of temperature is more apparent in $\Delta R_{xx}$ after subtracting the background from $R_{xx}$, (Fig. 2(b)). SdH oscillations are a periodic function of $1/B$. Evidence of multiple subbands in AlInN/GaN or AlGaN/GaN heterostructures has been reported [1,5]. However, Fourier Transform analysis of our data for $\Delta R_{xx}$ results in a single peak frequency $B_F = 403$ T as shown in the inset of Fig. 2(b). This indicates that only one band is dominant in our sample. Also the frequency $B_F$ is directly related to the carrier density by $n_{SdH} = 2eB_F/h$. Hence the carrier density of this sample is $n_{SdH} = 1.948 \times 10^{13} cm^{-2}$. This agrees very well with the averaged value obtained from our Hall measurement.

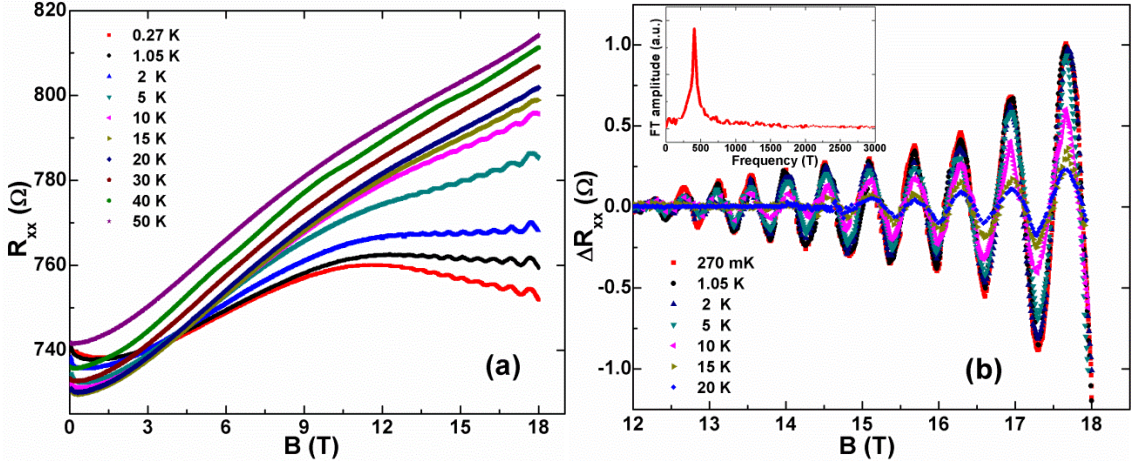

FIG. 2. (a) Magnetoresistance up to 18 T at a set of temperatures. (b) Shubnikov-de Haas oscillations after subtracting the background.

The amplitude of the SdH oscillation is given by [7,12] $\Delta R_{xx} = 4R_0 \frac{\chi}{\sinh(\chi)} \exp(\frac{-\pi}{\omega_c \tau_q})$. Here $\tau_q$ is the quantum lifetime, $\chi = 2\pi^2 k_B T/\Delta E$ and the Landau level energy gap is $\Delta E = \hbar \omega_c = \hbar eB/m^*$. $k_B$ is the Boltzmann constant, $\hbar$ is the reduced Planck constant, $e$ is the electron charge.

The effective mass $m^*$ of electrons can be extracted from the temperature dependence of the SdH amplitude at a constant magnetic field by the following ratio [13]

$$\frac{\Delta R_{xx}(T,B)}{\Delta R_{xx}(T_0,B)} = \frac{T \sinh(\chi(T_0))}{T_0 \sinh(\chi(T))} = \frac{T \sinh(2\pi^2 k_B T_0/\Delta E(B))}{T_0 \sinh(2\pi^2 k_B T/\Delta E(B))}, \qquad (1)$$

Here we chose the lowest temperature 0.27 K as $T_0$. Figure 3(a) shows the above ratio of amplitude at $T_0 = 0.27$ K and $B = 17.7$ T. Analyzing our data using Eq. (1) we can extract $\Delta E(B)$. The inset of Fig. 3(a) shows the field dependence of Landau level energy gap. Thus the corresponding effective mass is $m^* = 0.2327 m_e$, similar to the values reported in AlInN/GaN heterostructures which are $0.22 m_e$ and $0.25 m_e$ [1,20] and in AlGaN/GaN systems which are $0.23 m_e$ and $0.24 m_e$ respectively [21,22].

The quantum lifetime is obtained from the slope of the Dingle plot, as shown in Fig. 3(b), because

$$ln\mathfrak{D} = \ln[\frac{\Delta R(T,B)\sinh(2\pi^2 k_B T/\Delta E(B))}{2\pi^2 k_B T/\Delta E(B)}] = C_0 - \frac{\pi m^*}{e\tau_q B}, \qquad (2)$$

where $\mathfrak{D}$ is the expression within the bracket, $C_0$ is a constant. In this sample the quantum lifetime is $\tau_q = 0.035\ ps$. Furthermore, $\tau_q$ also determines the Dingle temperature $T_D = h/(4\pi^2 k_B \tau_q)$, which is a measure of the disorder. At $T = 0.27$ K, we find a relatively high value $T_D = 34.7$ K. Also the broadening of the Landau levels [7,12], as determined by $k_B T_D \sim 3$ $meV$, is not much smaller than the Landau level spacing $\Delta E(B) = 9.04\ meV$ at 17.7 T. This may explain the relatively small amplitude ($\Delta R_{xx}/R_{xx} \ll 1$) of the SdH oscillations.

It is instructive to compare the quantum relaxation rate to transport rate, since $1/\tau_q = \int P(\theta)d\theta$ and $1/\tau_t = \int P(\theta)(1-\cos\theta)d\theta$, where $P(\theta)$ is the probability of scattering through an angle $\theta$. The quantum lifetime $\tau_q$ includes information of all scatterings; however the transport lifetime $\tau_t$ obtained from Hall mobility is weighted by the scattering angle and mainly determined by large angle scattering [21,23,24]. Here the transport lifetime $\tau_t = 0.252\ ps$ is nearly an order of magnitude larger than the quantum lifetime. In particular the ratio $\tau_t/\tau_q = 7.2$ indicates that small angle scattering associated with long range interaction due to distant ionized impurities is the dominant scattering mechanism in our sample.

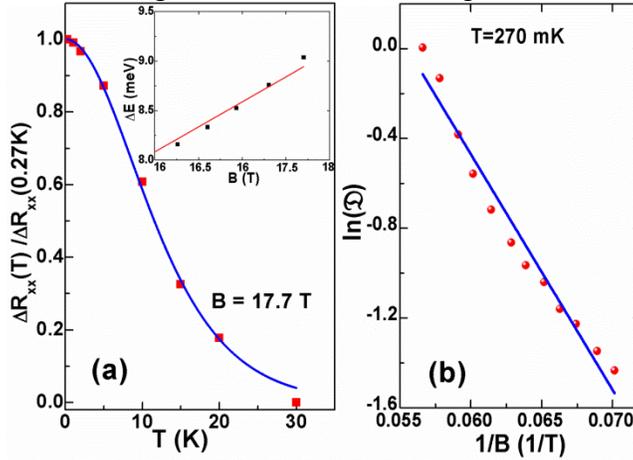

FIG. 3. (a) Effective mass plot at 17.7 T, where the data are best fit to Eq. (2); Inset is field dependence of the Landau level spacing. (b) Dingle plot to obtain the quantum lifetime in AlInN/GaN heterostructure.

## C. Weak localization

As is evident from Fig. 2(a), at low magnetic fields $R_{xx}$ decreases with the applied field; that is conductivity goes up with increasing field. This negative magnetoresistance arises from weak localization. It is convenient to define the magnetoconductivity by $\sigma_{xx} = \rho_{xx}/(\rho_{xx}^2 + \rho_{xy}^2)$. The quantum correction to the change in magnetoconductivity at low magnetic fields is [5,22,25,26]

$$\Delta\sigma_{xx} = \sigma_{xx}(B) - \sigma_{xx}(0) = \frac{e^2}{\pi h}[\psi\left(\frac{1}{2} + \frac{\hbar}{4eDB\tau_i}\right) - \psi\left(\frac{1}{2} + \frac{\hbar}{4eDB\tau_e}\right) + \ln\left(\frac{\tau_i}{\tau_e}\right)], \qquad (3)$$

Here $\psi$ is the digamma function; $\tau_i$ and $\tau_e$ are the inelastic and elastic scattering times respectively; $D$ is the diffusion constant given by $D = v_F^2 \tau/2$, and for our two-dimensional system the Fermi velocity $v_F = \hbar k_F/m^* = \hbar\sqrt{2\pi n}/m^* = 0.5504 \times 10^6 m/s$. By choosing parameter value estimated earlier $\tau \equiv \tau_t = 0.252\ ps$, we determined $D = 0.03817 m^2/s$ and the mean free path $l = v_F\tau_t = 139\ nm$ [6].

The inelastic and elastic scattering times were computed from the best fit analysis of experimental data to Eq. (3). $\Delta\sigma_{xx}$ for a set of temperatures is shown in Fig. 4(a). For a constant temperature the conductivity increases with increasing magnetic field, and at higher temperature, the conductivity is reduced. We find that Eq. (3) describes the experimental data very well for temperatures below 20 K, which allows us to obtain the values of the relaxation times. Interestingly the elastic scattering time $\tau_e$ is constant with temperature; this may be due to the short range interactions such as impurity or interface roughness scatterings. Also $\tau_e = 0.144 ps$ is the same order as the transport time $\tau_t$. However, the inelastic scattering time (phase coherence time) $\tau_i$ is much larger than the elastic scattering time and transport lifetime at low temperatures. This is necessary for the observation of weak localization, because the phase coherence length should be long enough so that the carrier can return to the origin after several scatterings. In addition, $\tau_i$ decreases with increasing temperature. In fact, the inelastic scattering rate is linearly proportional to temperature, $\tau_i^{-1} \propto T$, as shown in Fig. 4(b). This linearity has been attributed to phase breaking by inelastic electron-electron scattering [4,5].

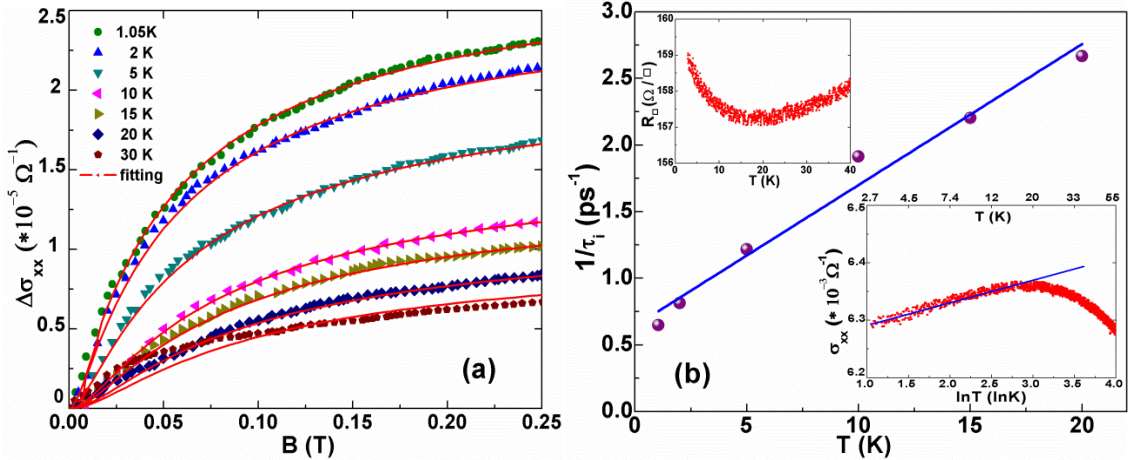

FIG. 4. (a) Magnetoconductivity at low magnetic fields for several temperatures. (b) The inelastic scattering rate displays linear temperature dependence. Insets are the zero-field resistance and conductivity respectively.

The effect of electron-electron scattering is also observed in the absence of the magnetic field. As shown in the inset of Fig. 4(b), the sheet resistance at low temperatures is non-monotonic. With increasing temperature it first decreases until 20 K and then increases. This enhanced conductivity for less than 20 K is also due to weak localization. Because the quantum interference correction at zero magnetic field is [25,27],

$$\sigma_{xx}(B=0) = \frac{e^2}{\pi h}\left[\psi\left(\frac{1}{2} + \frac{\hbar}{4eDB\tau_i}\right) - \psi\left(\frac{1}{2} + \frac{\hbar}{4eDB\tau_e}\right)\right] \cong -\frac{e^2}{\pi h}\ln\frac{\tau_i}{\tau_e}, \qquad (4)$$

since $\psi(x) \to \ln x$ when $x \gg 1$.

As stated above, $\tau_i \propto T^{-1}$, thus $\sigma_{xx}(0) \propto \ln T$ [28]. The conductivity at zero-field is plotted as a function of $\ln T$ in the inset of Fig. 4(b). Clearly below 20 K, $\sigma_{xx}(B=0)$ displays a linear dependence, characteristic of weak localization. Hence at low temperatures, we observed the experimental evidence of weak localization in zero-field transport as well as the magnetotransport in our AlInN/GaN system, both hallmarks of electron-electron scattering.

## D. Angle dependence

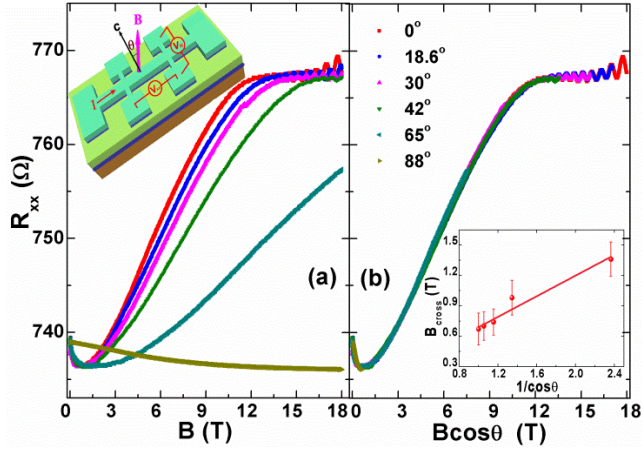

FIG. 5. (a) Angle dependence of the magnetoresistance at 2 K. (b) The magnetoresistance as a function of the perpendicular field, all data collapse to a single curve.

To further investigate weak localization behavior, we varied the angle $\theta$ between the applied magnetic field and c axis of crystal. Figure 5 shows the angle-dependent resistance at T = 2 K. At 0°, when magnetic field is perpendicular to the sample, the magnetoresistance is pronounced and similar to the behavior described earlier. With the increase of tilt angle, the influence of the magnetic field becomes smaller and weak localization is maintained over higher magnetic fields. At the highest tilt angle ($\theta = 88°$), effects of weak localization dominate the entire field regime such that the resistance continues to decrease with increasing magnetic field, displaying a negative magnetoresistance even up to 18 T. The crossover field where the magnetoresistance is lowest displays the anticipated linear dependence on $1/cos\theta$. Furthermore, the SdH oscillations disappear gradually with increasing angle. If we plot the magnetoresistance as a function of the perpendicular component of the magnetic field, $Bcos\theta$, the peaks collapse respectively for different angles, as shown in Fig. 5(b). Therefore, the behavior is controlled only by the perpendicular magnetic field, confirming the two dimensional nature of electron transport in this heterostructure [6,12,13].

## E. Discussion

The sample discussed earlier, labeled as sample A, shows small amplitude Shubnikov-de Haas oscillations. In comparison, sample B, deposited under different conditions with slightly lower carrier density $n_{SdH} = 1.53 \times 10^{13} cm^{-2}$, has a larger amplitude SdH oscillation. As shown in Fig. 6(a), the maximum of $\Delta R_{xx}$ for sample B is around 10 $\Omega$, one order of magnitude bigger than that of sample A. In specimen B the quantum lifetime is $\tau_q = 0.074\ ps$, which is also smaller than the transport lifetime. The ratio $\tau_t/\tau_q = 6.1$, although slightly less than the ratio in sample A, still indicates that small angle scattering is the dominant scattering mechanism. Moreover, the change of conductivity $\Delta\sigma$ due to weak localization in sample B is larger than that in sample A (Fig. 6(b)). Hence the interference is much stronger. The analysis of Eq. (3) shows that the inelastic scattering time of sample B is correspondingly bigger than that of sample A.

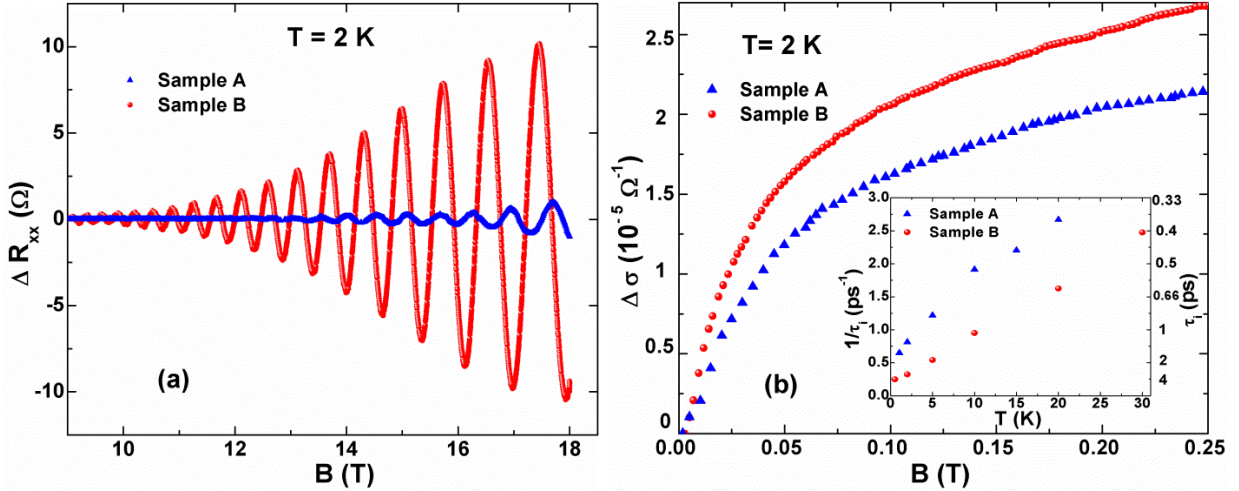

FIG. 6. (a) Shubnikov-de Haas oscillations for two samples. Clearly, Sample B has stronger SdH oscillations. (b) weak licalization of two samples at 2 K. Inset is the inelastic relaxiation time as a function of temeprature.

TABLE I. Comparison of Shubnikov-de Haas oscillation and weak localization parameters in GaN based 2DEG. The inelastic scattering time is the value at lowest temperature respectively.

| System | Carrier density $(10^{13} cm^{-2})$ | Quantum lifetime $\tau_q (ps)$ | Transport lifetime $\tau_t (ps)$ | Inelastic time $\tau_i (ps)$ |
| --- | --- | --- | --- | --- |
| $Al_{0.83}In_{0.17}N/GaN$ (A) | 1.94 | 0.035 | 0.252 | 1.66 |
| $Al_{0.83}In_{0.17}N/GaN$ (B) | 1.53 | 0.074 | 0.45 | 4.02 |
| $Al_{0.25}Ga_{0.75}N/GaN$ (ref.[22]) | 1.01 | 0.050 | 0.26 | 4.00 |
| $Al_{0.22}Ga_{0.78}N/GaN$ (ref.[29]) | 1.25 | 0.078 | 0.13 | 4.67 |

Table I lists the carrier density, quantum ($\tau_q$), transport ($\tau_t$) lifetimes and inelastic scattering (phase coherence) time $\tau_i$ in related systems. Qualitatively, the inelastic scattering rate should be smaller than transport rate, which in turn is less than the quantum relaxation rate ($\frac{1}{\tau_i} < \frac{1}{\tau_t} < \frac{1}{\tau_q}$). Indeed, for both samples A and B, $\tau_q$ is the shortest, and about an order of magnitude smaller than $\tau_t$, as expected when small angle scattering is dominant. Furthermore, the inelastic time scale is the longest, and much longer than quantum and transport lifetimes. This same trend in time scales is also reported in AlGaN/GaN systems, as can be seen in table I.

Physically, the SdH effect is related to the momentum space ($k$ space), it arises from the tuning of the density of states as well as the Fermi level by the magnetic field. With increasing magnetic field, Landau levels periodically cross the Fermi level, resulting in oscillations of conductance. Whereas, weak localization is associated to non-classical back scattering of the carrier to the origin, where small angle scattering by long range interaction is ineffective for this process [10]. As a matter of fact, weak localization is an interference phenomenon hence sensitive to the phase of the wavefunction. At low temperatures, although small-angle scattering is frequent, only interactions that destroy phase coherence such as electron-electron and electron-phonon scatterings play important roles in the weak localization process [30].

## IV. CONCLUSION

Weak localization is observed for the first time in AlInN/GaN heterostructures at low temperatures (T < 20 K), which indicates a long phase coherence length. The zero-field conductivity varies as *lnT* and the magnetoconductivity increases with increasing magnetic field for low fields; both of which are hallmarks of weak localization. Electron-electron scattering is the dominant phase breaking mechanism in this temperature range. At high magnetic fields, the resistance below 20 K exhibits Shubnikov-de Haas oscillations. The large ratio between the transport and quantum lifetimes indicates small angle scattering is dominant at low temperatures. Our angle dependent measurement shows that the magnetoresistance scales with *Bcosθ*, confirming the two dimensionality of system. Below 20 K the carrier mobility is weakly reduced with increasing temperature. In contrast, above 20 K the mobility rapidly decreases with increasing temperature as the source of scattering changes from acoustic phonons to optical phonons.

In summary, despite the long phase coherence time at low temperatures in the 2DEG of AlInN/GaN heterostructures, electron-electron scattering, small angle scattering due to long-range Coulomb interactions, acoustic phonons and optical phonons all progressively contribute to the decrease in mobility with rising temperature. In order to improve the performance of high power, high frequency electronic devices and high electron mobility transistors, these scatterings need to be suppressed.

## ACKNOWLEDGEMENTS

We thank David Weber for the help of transport measurements, Ning Lu for help in the sample fabrication. The work is supported in part by DOE award DE-NA0002630 and University of South Carolina. A portion of this work was performed at the National High Magnetic Field Laboratory, which is supported by NSF Cooperative Agreement No. DMR-1157490 and the State of Florida.We thank David Weber for the help of transport measurements, Ning Lu for help in the sample fabrication. The work is supported in part by DOE award DE-NA0002630 and University of South Carolina. A portion of this work was performed at the National High Magnetic Field Laboratory, which is supported by NSF Cooperative Agreement No. DMR-1157490 and the State of Florida.